\documentclass[prb,showpacs,twocolumn,superscriptaddress]{revtex4-1}

\usepackage{bm}
\usepackage{amsmath}
\usepackage{amsfonts}
\usepackage[pdftex]{graphicx}
\usepackage[pdftex]{epsfig}
\usepackage{epstopdf}
\usepackage{color}

\newcommand{\beq}{\begin{equation}}
\newcommand{\eeq}{\end{equation}}
\newcommand{\beqa}{\begin{eqnarray}}
\newcommand{\eeqa}{\end{eqnarray}}
\newcommand{\vep}{\varepsilon}
\newcommand{\rd}{|\negthickspace \Downarrow \rangle}
\newcommand{\ld}{\langle \Downarrow \negthickspace |}
\newcommand{\ru}{|\negthickspace \Uparrow \rangle}
\newcommand{\lu}{\langle \Uparrow \negthickspace |}

\begin{document}

\title{Probing Quantum Phase Transitions on a Spin Chain with a Double Quantum Dot}
 \author{Yun-Pil Shim}
 \altaffiliation{Current address: Laboratory for Physical Sciences, College Park, Maryland 20740, USA}
 \affiliation{Department of Physics, University of Wisconsin-Madison, Madison, Wisconsin 53706, USA}
 \author{Sangchul Oh}
 \affiliation{Department of Physics, University at Buffalo, State University of New York, Buffalo, New York 14260, USA}
 \author{Jianjia Fei}
 \affiliation{Department of Physics, University of Wisconsin-Madison, Madison, Wisconsin 53706, USA}
 \author{Xuedong Hu}
 \affiliation{Department of Physics, University at Buffalo, State University of New York, Buffalo, New York 14260, USA}
 \author{Mark Friesen}
 \affiliation{Department of Physics, University of Wisconsin-Madison, Madison, Wisconsin 53706, USA}
 \date{\today}

\begin{abstract}
Quantum phase transitions (QPTs) in qubit systems are known to produce singularities in the entanglement, which could in turn be used to probe the QPT. 
Recent proposals have suggested that the QPT in a spin chain could be probed via the entanglement between external qubits coupled to the spin chain. 
Such experiments may be technically challenging, because the probe qubits are nonlocal. 
Here we show that a double quantum dot coupled locally to a spin chain provides an alternative and efficient probe of a QPT. 
To demonstrate this method in a simple geometry, we propose an experiment to observe a QPT in a triple quantum dot, based on the well-known singlet projection technique.  
\end{abstract}

\pacs{03.67.Lx, 73.21.La, 75.10.Pq, 64.70.Tg, 05.30.Rt, 85.35.Gv}
\maketitle

\section{Introduction}\label{sec:1_introduction} 
Spin chains have been studied for many years because of their simple formulation, which enables analytical solutions, and their similarity to more complex quantum many-body systems.
Recently, it has been possible to engineer fully tunable spin chains based on quantum dots with one or more electrons.\cite{jacak_hawrylak_book1998}
Such chains can be used as spin qubits for quantum computing,~\cite{loss_divincenzo_pra1998}
as a spin bus whose ground state transmits quantum information over large distances,~\cite{li_shi_pra2005,oh_wu_pra2011} 
or to mediate interactions between nonlocal  qubits.~\cite{venuti_boschi_prl2006,friesen_biswas_prl2007,oh_prb2010,shim_oh_prl2011,fei_zhou_pra2012} 
In all these settings, the ability to mediate entanglement is paramount for incorporating spin chains into quantum devices.

Quantum phase transitions (QPTs) can have a strong effect on adiabatic operations involving the ground state of a spin chain.
QPTs occur at energy level crossings as a function of  external parameters, between ground states with very different physical properties.~\cite{sachdev_book} 
In the critical regime where the two ground states are nearly degenerate, macroscopic observables such as two-qubit entanglement can exhibit non-analytic behavior.~\cite{osterloch_amico_nature2002,osborne_nielsen_pra2002,vidal_latorre_prl2003,%
gu_lin_pra2003,tagliacozzo_oliveira_prb2008,pollmann_mukerjee_prl2009}
In finite-size systems, the phase transition is typically discontinuous, or first order; however, the underlying physics is essentially the same as in infinite systems.

From a quantum information perspective, QPTs may produce singularities in the entanglement, which could potentially enhance device operation.  
Alternately, we could view entanglement as a sensitive probe of the ground state, which could potentially enhance our understanding of QPTs. 
The singularities in question reflect the entanglement properties between the constituent spins in the chain, or between external qubits that are weakly coupled to the system.
The latter approach was recently applied to $XY$-type spin chains, using nonlocal pairs of qubits to probe the QPT.~\cite{yi_cui_pra2006,yuan_zhang_pra2007b,ai_shi_pra2008}
For this arrangement, strong enhancements of the entanglement between the probe qubits were observed near the critical points of the spin chain.
In realistic quantum dots however,  the spin couplings are not of the $XY$ type, and nonlocal measurements can be rather challenging.~\cite{shulman_science2012}

Because of its simplicity, a local probe could potentially be more effective.
Unfortunately, the simplest type of measurement -- mapping out the effective interaction between the spin chain and a weakly coupled qubit -- does not exhibit unusual behavior near an energy level crossing.~\cite{shim_oh_prl2011}
It has been suggested that the time evolution of a coupled qubit could be used to probe a QPT~\cite{quan_song_prl2006,yuan_zhang_pra2007a} via the Loshmidt echo.~\cite{gorin_prosen_physrep2006} 
This method has been successfully implemented using nuclear magnetic resonance (NMR). \cite{zhang_peng_prl2008,zhang_cucchietti_pra2009} 
Such dynamical methods are powerful, but they are still more challenging than simple projective measurements.

In this paper, we propose a local, projective scheme for detecting QPTs in a quantum dot spin chain.
The chain undergoes successive QPTs as a function of the external magnetic field, 
as is described in Sec.~\ref{sec:2_energy_levels}.
We study two different external probes of the QPT. 
First, in Sec.~\ref{sec:3_nonlocal}, we consider a nonlocal probe, consisting of two qubits weakly coupled to the chain at different locations, as depicted in Fig.~\ref{fig:1_system}(a). 
We calculate the entanglement between the probe qubits numerically, using the concurrence measure $C$,~\cite{wootters_prl1998} and we observe singularities when the spin chain undergoes a QPT.
We also obtain analytical estimates for the entanglement using perturbation theory.
Next, in Sec.~\ref{sec:4_local}, we consider a local probe, consisting of a double quantum dot coupled to a single node of the spin chain, as shown in Fig.~\ref{fig:1_system}(d).
We find that the ground state properties of the chain are imprinted onto the probe, and we investigate the concurrence singularities both numerically and analytically.
Interestingly, we find that the probability $P_S$ for the probe qubits to form a singlet state echos the non-analytic response of the concurrence.
This is significant because the singlet probability is relatively easy to measure, using the singlet projection technique common to spin qubit experiments.~\cite{Petta2005} 
In Sec.~\ref{sec:5_experiment} we propose a simple experiment to test these concepts on the smallest possible spin system of size $N=1$.
This ``chain" has a single energy level crossing as a function of magnetic field, and the ground state transition exhibits a non-analyticity consistent with a QPT.
Our proposal involves a total of three quantum dots, and it is therefore within reach of current triple dot technologies.~\cite{schroer_greentree_prb2007,amaha_hatano_prb2012,gaudreau_granger_natphys2012,hsieh_shim_rpp2012}
A brief summary and conclusions are given in Sec.~\ref{sec:6_summary}.

\section{Energy level crossings in the spin chain}\label{sec:2_energy_levels}
\begin{figure}
  \includegraphics[width=\linewidth]{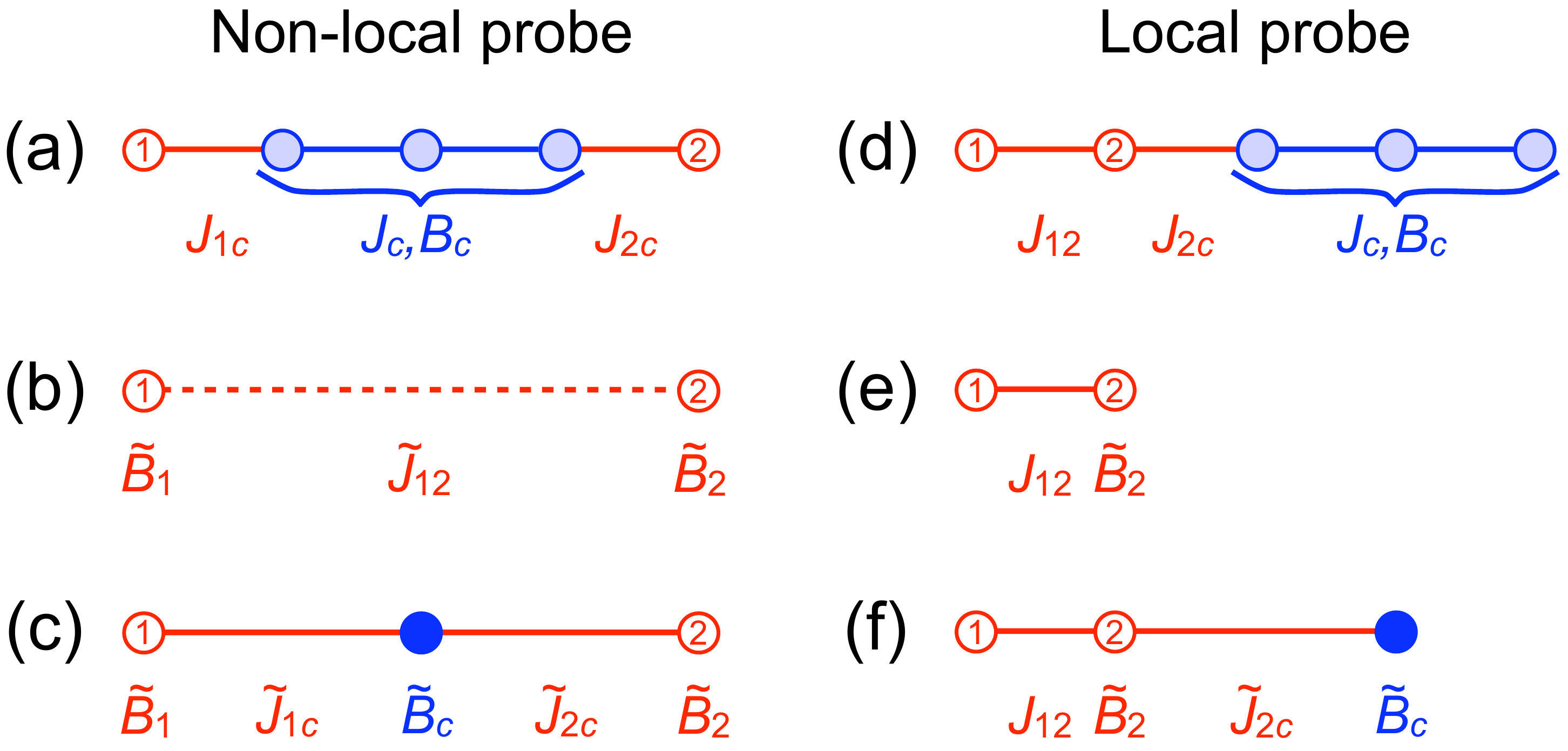}\\
  \caption{(color online) Graphical representations of the Hamiltonians studied here.
  (a)-(c) correspond to a nonlocal probe in which the probe qubits (1 and 2) are attached to different nodes on the spin chain.
  (d)-(f) correspond to a local probe in which a pair of qubits is attached to a single node on the chain.
  (a) and (d) describe the full, physical geometry, where the chain (lightly shaded circles) is formed of an arbitrary number of physical spins.
  Here, we show probe qubits attached to the endpoints of the chain; however, similar results are obtained for any attachment points.
  (b), (c), (e), and (f) represent effective geometries, in which the spin chain in its ground state is replaced by a pseudospin, as indicated by a filled circle (when appropriate).
  (b) and (e) represent noncritical Hamiltonians (far away from a critical point), where the bus ground state is non-degenerate (pseudospin-0).
  (c) and (f) represent critical Hamiltonians, where the bus ground state is doubly-degenerate (pseudospin-1/2).
  In the effective Hamiltonians, the bus pseudospin interacts with the probe qubits via effective couplings ($\widetilde{J}$) and effective fields ($\widetilde{B}$).
  In (b), the effective coupling $\widetilde{J}_{12}$ is weak (\emph{i.e.}, second order), as indicated by a dashed line.  }
  \label{fig:1_system}
\end{figure}

We adopt an isotropic Heisenberg model of a spin chain, as appropriate for single-electron spins in quantum dots.
The Hamiltonian for a chain of length $N$ is given by 
\begin{equation}\label{eq:Hspinchain}
H_c = J_c \sum_{j=1}^{N-1} \mathbf{s}_j \cdot \mathbf{s}_{j+1} - B_c \sum_{j=1}^{N} s_{jz}
\end{equation}
where $\mathbf{s}_j$ are spin operators for the individual electrons in the chain. 
The bare exchange couplings between the spins are labeled $J_c$, and the applied magnetic field is $\mathbf{B}_c=B_c\hat{\bf z}$.
Throughout this paper, we will adopt $J_c$ as the unit of energy.
Magnetic fields will also be expressed in energy units.
A graphical representation of the chain Hamiltonian is given by the lightly shaded circles in Figs.~\ref{fig:1_system}(a) and (d).

For now, we ignore any couplings to external qubits, and calculate the energy spectrum for $H_c$ as a function of $B_c$.
The results are shown in Figs.~\ref{fig:2_Energy_Entanglement}(a) and (b) for the cases $N=4$ and 5.
For the uniform field we consider here, energy levels are straight lines as functions of the external field due to the Zeeman energy.
In this paper, we are most interested in the two lowest energy levels for a given value of $B_c$, whose crossings are indicated by circles in Fig.~\ref{fig:2_Energy_Entanglement}.
Each level crossing is associated with a QPT in the finite-size spin chain, and the $z$ component of the total spin changes by 1 at each critical point.

\section{Nonlocal probes}\label{sec:3_nonlocal}
We now couple the spin chain to two external qubits, labelled 1 and 2 in Fig.~\ref{fig:1_system}, which will serve as probes of the QPT.
In this section, we consider only the nonlocal probe geometry shown in Fig.~\ref{fig:1_system}(a),
with the coupling Hamiltonian given by
\begin{equation}\label{eq:H_nl_probe}
H_p= J_{1c} \mathbf{S}_1 \cdot \mathbf{s}_{1} + J_{2c} \mathbf{S}_2 \cdot \mathbf{s}_{N} .
\end{equation}
Here, $\mathbf{S}_1$ and $\mathbf{S}_2$ are the spin operators for the probe qubits.
The probes may be coupled to any node of the spin chain, with similar results.
For definiteness here, we have attached them to the endpoints of the chain.

\begin{figure}
  \includegraphics[width=\linewidth]{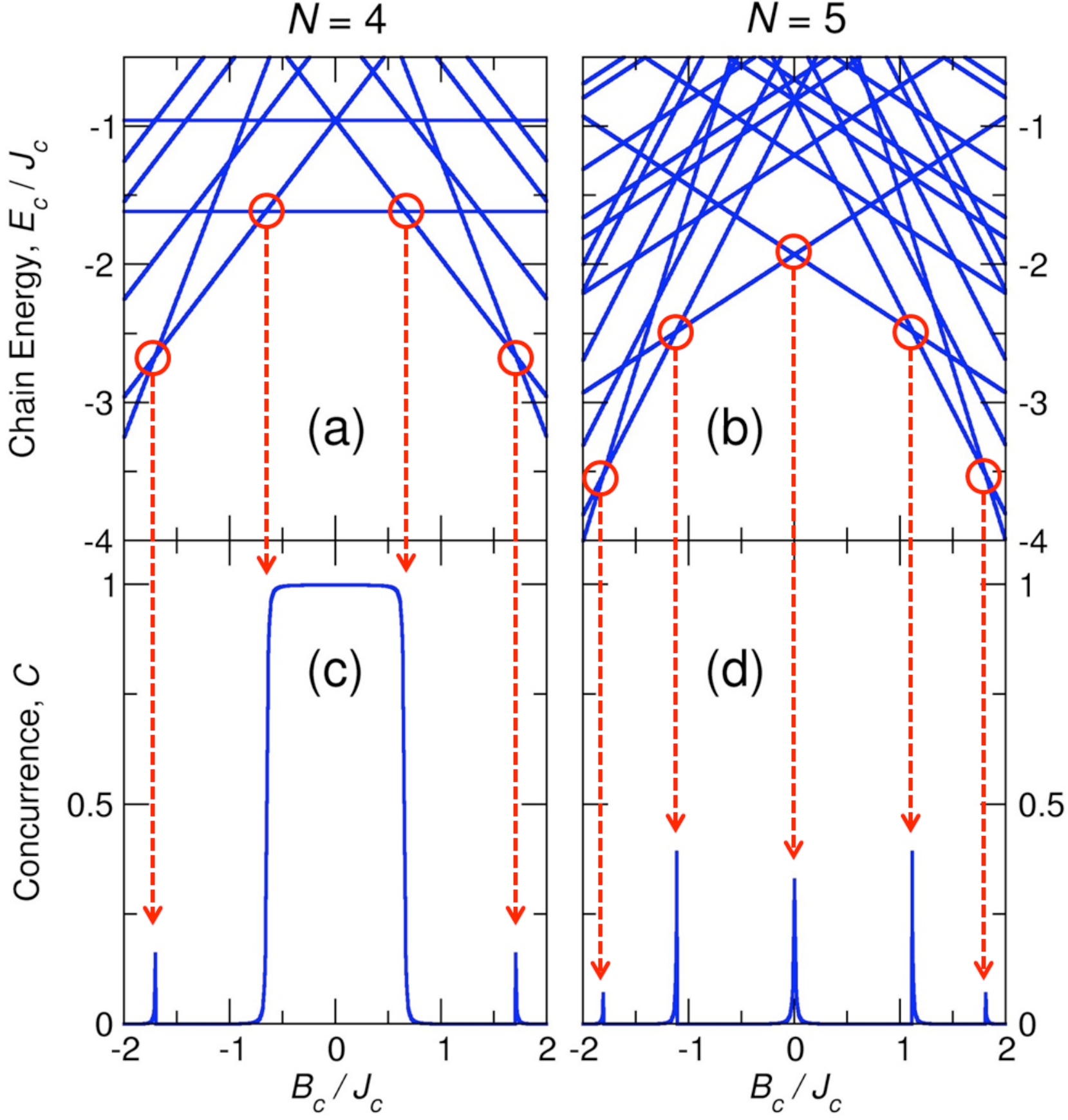}\\
  \caption{(color online) (a), (b) Energy spectra of spin chains of length $N=4$ and 5, with no coupled qubits, as a function of the magnetic field $B_c$ expressed in energy units, and scaled by the coupling constant $J_c$.
  The energy level crossings of the ground state are indicated by circles.
  (c), (d) The corresponding concurrence between the probe qubits, when they are coupled in the nonlocal probe geometry of Fig.~\ref{fig:1_system}(a).
  Here, $C$ is dimensionless, 
 and we take the coupling between the probe qubits and the spin chain to be $J_{1c}$=$J_{2c}$=$0.02J_c$.  (See Fig.~\ref{fig:1_system}.)
  The singular features of the concurrence occur at energy level crossings of the chain.
          }
  \label{fig:2_Energy_Entanglement}
\end{figure}

When the probe couplings are turned on, the energy levels of the chain expand into energy manifolds.
Our goal is to probe the QPT without disturbing it, so the manifold structure in Figs.~\ref{fig:2_Energy_Entanglement}(a) and (b) should remain largely undisturbed.
This places constraints on the probe couplings.
First, the bare coupling constants must be small, such that $J_{1c},J_{2c}\ll J_c$.
The magnetic field applied to the probe qubits should also be much smaller than $J_c$, necessitating a magnetic field gradient between the qubits and the chain.
For definiteness, we take the magnetic field on the probe qubits to be zero.
Although large field gradients are difficult to achieve in the laboratory, we will focus on QPTs occurring at zero field, in an odd-size chain.
For this case, the field gradient is small, and does not pose a serious experimental challenge.

The concurrence calculation is performed after first tracing out the spin-chain degrees of freedom from the full Hamiltonian, $H=H_c+H_p$, to obtain the reduced, bipartite density matrix for the probe qubits, $\rho_{12}$. 
The concurrence is defined as~\cite{wootters_prl1998} 
\begin{equation}
C=\max \{ 0,\sqrt{\vep_1}-\sqrt{\vep_2}-\sqrt{\vep_3}-\sqrt{\vep_4} \} , 
\label{eq:concur}
\end{equation}
where $\vep_1 \ge \vep_2 \ge \vep_3 \ge \vep_4 \ge 0$ are the eigenvalues of the operator $R=\rho_{12} \left( \sigma_y \otimes \sigma_y \right) \rho_{12}^* \left( \sigma_y \otimes \sigma_y \right)$, and $\sigma_y$ is a Pauli spin matrix.
Figures \ref{fig:2_Energy_Entanglement}(c) and (d) show numerical results for the concurrence between the probe qubits, in the nonlocal coupling geometry.
We see that the concurrence exhibits singularities which are correlated with the energy level crossings of the spin chain, as expected for QPTs.
Away from the level crossings, the concurrence falls quickly to zero.  
The only exception to this regular behavior is observed near zero field for even-size chains, where the concurrence plateaus at its maximum value, $C=1$.

This interesting behavior can be understood intuitively by treating the probe qubits as a perturbation.
It has previously been shown that when the bare qubit-chain coupling is small, and when the chain is in its ground state, the interactions can be described by effective Hamiltonians.
(These results were first obtained in [\onlinecite{shim_oh_prl2011}]. Analytical expressions for the effective Hamiltonians, and a brief summary of the results, are provided in Appendix~\ref{sec:A_eff_param}.)
Far away from a QPT, the system is noncritical and the effective Hamiltonian involves only the external spins:
\begin{equation}\label{eq:H_nl_nc}
\widetilde{H}_\text{nc} =  \widetilde{B}_1 S_{1z} + \widetilde{B}_2 S_{2z} + \sum_{\alpha =x,y,z}
\widetilde{J}_{12\alpha} S_{1\alpha} S_{2\alpha}~,
\end{equation}
as indicated graphically in Fig.~\ref{fig:1_system}(b).
Here, the bare coupling parameters $J_{1c}$ and $J_{2c}$ are hidden inside the effective coupling $\widetilde{J}_{12}$ and the effective local fields $\widetilde{B}_{1,2}$.
We note that the effective couplings are generally anisotropic, except in special cases, since the external field breaks the spin rotational symmetry.~\cite{shim_oh_prl2011}
If the qubit couplings are turned on adiabatically, the chain will remain in an inert, effective pseudospin-0 state.
The effective coupling arises due to virtual excitations of the chain outside its ground-state manifold; it is therefore second order in the perturbation:  $\widetilde{J}_{12}/J_c \sim (J_{1c}/J_c)^2$.
On the other hand, the effective fields emerge at first order:  $\widetilde{B}_{1,2}/J_c \sim (J_{1c,2c}/J_c)^1$.
We therefore generally find that $\widetilde{B}_{1,2}\gg \widetilde{J}_{12}$ in the noncritical regime.
Hence, the external qubits align with the effective field to form a separable state, for which $C\simeq 0$.
The only exception is the special case near $B_c=0$ for an even-size chain.
Here $\widetilde{B}_{1,2}=0$ due to the spin-singlet character of the chain ground state.
Since $\widetilde{J}_{12}\neq 0$, it can generate maximal entanglement between the probe qubits, as shown in Fig.~\ref{fig:2_Energy_Entanglement}(c).

The situation is very different near a QPT.
In this case, the ground state of the chain is approximately two-fold degenerate and behaves as an effective pseudospin-1/2, as indicated in Fig.~\ref{fig:1_system}(c).
The effective Hamiltonian at the critical point then describes a simple three-body system:
\begin{eqnarray}\label{eq:H_nl_c}
\widetilde{H}_\text{cp} &=& \widetilde{B}_1 S_{1z} + \widetilde{B}_2 S_{2z} - \widetilde{B}_c S_{cz} \nonumber\\
&& + \sum_{\alpha =x,y,z} \left( \widetilde{J}_{1c\alpha} S_{1\alpha} S_{c\alpha}
                              +\widetilde{J}_{2c\alpha} S_{2\alpha} S_{c\alpha} \right)~.
\end{eqnarray}
Here, the spin operator $\mathbf{S}_c$ acts on the the pseudospin-1/2 of the chain ground state. 
In contrast with the noncritical regime, the effective couplings are now first order in the perturbation:  $\widetilde{J}_{1c,2c}/J_c \sim J_{1c,2c}/J_c$.
Because $\widetilde{J}_{1c,2c}$ is relatively large, $\widetilde{H}_\text{cp}$ can mediate the entanglement between the two probe qubits, with the resulting value of the concurrence determined by the relative size of $\widetilde{J}_{1c,2c}$ compared to $\widetilde{B}_{1,2}$.
The couplings $\widetilde{J}_{1c,2c}$ enhance entanglement while the fields $\widetilde{B}_{1,2}$ suppress it.

We can take this analysis further for the QPT occurring at $B_c=0$ in an odd-size spin chain [\emph{e.g.}, the central peak in Fig.~\ref{fig:2_Energy_Entanglement}~(d)].
In this case,  the effective Hamiltonian has a much simpler, isotropic form, with $\widetilde{J}_{1c}$=$\widetilde{J}_{2c}$=$\widetilde{J}$, $\widetilde{B}_{1,2}$=$0$, $\widetilde{B}_c$=$B_c$, and $\widetilde{H}_\text{cp} = \widetilde{J} (\mathbf{S}_1+\mathbf{S}_2)\cdot\mathbf{S}_c - B_c S_{c,z}$.
We can derive an expression for the concurrence between the probe qubits which shows the explicit form of the non-analyticity.
Without loss of generality, we will only consider the case $\widetilde{B}_c \ge 0$.

We must first determine the ground state of the full system.
Since the spin operators $S_{\text{tot},z}=S_{1,z}+S_{2,z}+S_{c,z}$ and $S_{12}^2=(\mathbf{S}_1+\mathbf{S}_2)^2$ commute with the Hamiltonian (\ref{eq:H_nl_c}), we can label the individual subspaces according to their quantum numbers.  
For the subspace with $S_{\text{tot},z}=\pm 3/2$ and $S_{12}=1$, the eigenstates are  $| S_{1,z}, S_{c,z}, S_{2,z} \rangle=|\!\uparrow \uparrow \uparrow \rangle$ and $|\! \downarrow \downarrow \downarrow \rangle$, respectively. 
The eigenenergies are $E_{\pm 3/2} = \widetilde{J}/2 \mp \widetilde{B}_c/2$. 
For $S_{\text{tot},z}=1/2$ and $S_{12}=0$, the eigenstate is $ ( |\!\uparrow\uparrow\downarrow\rangle - |\!\downarrow\uparrow\uparrow\rangle )/\sqrt{2}$ and the eigenenergy is $-\widetilde{B}_c/2$. 
For $S_{\text{tot},z}=1/2$ and $S_{12}=1$, the subspace is two-dimensional; in the ordered basis  
$\{ |\!\uparrow\downarrow\uparrow\rangle, ( |\!\uparrow\uparrow\downarrow\rangle + |\!\downarrow\uparrow\uparrow\rangle )/\sqrt{2} \}$ we have
\begin{equation}
H_{1/2} = \left( \begin{array}{cc}
           \frac{\widetilde{B}_c-\widetilde{J}}{2}  & \frac{\sqrt{2}}{2} \widetilde{J} \\
           \frac{\sqrt{2}}{2} \widetilde{J} & -\frac{\widetilde{B}_c}{2}
           \end{array}
    \right)~.
\end{equation}
The lowest eignevalue in this subspace is $E_{1/2}= -\widetilde{J}/4 - A_{1/2}/2$, where $A_{1/2}=\sqrt{\frac{9}{4}\widetilde{J}^2 + \widetilde{B}_c^2 - \widetilde{J} \widetilde{B}_c}$.
For $S_{\text{tot},z}=-1/2$ and $S_{12}=0$, the eigenstate is $ ( |\!\uparrow\downarrow\downarrow\rangle - |\!\downarrow\downarrow\uparrow\rangle )/\sqrt{2}$ and the eigenenergy is $\widetilde{B}_c/2$. 
For $S_{\text{tot},z}=-1/2$ and $S_{12}=1$, the subspace is two-dimensional; in the ordered basis 
$\{ ( |\!\uparrow\downarrow\downarrow\rangle + |\!\downarrow\downarrow\uparrow\rangle )/\sqrt{2} , |\!\downarrow\uparrow\downarrow\rangle,\}$ we have
\begin{equation}
H_{-1/2} = \left( \begin{array}{cc}
           \frac{\widetilde{B}_c}{2} & \frac{\sqrt{2}}{2} \widetilde{J} \\
           \frac{\sqrt{2}}{2} \widetilde{J} & -\frac{\widetilde{B}_c+\widetilde{J}}{2}
           \end{array}
    \right)~.
\end{equation}
The lowest eignevalue in this subspace is $E_{-1/2}= -\widetilde{J}/4 - A_{-1/2}/2$, where $A_{-1/2}=\sqrt{\frac{9}{4}\widetilde{J}^2 + \widetilde{B}_c^2 + \widetilde{J} \widetilde{B}_c}$.
When $\widetilde{B}_c \ge 0$, we can easily see that $E_{-1/2}$ is the ground state energy, and its eigenstate  is given by
\begin{equation}
|\Psi_\text{GS}\rangle = \frac{1}{\sqrt{2}}
           \left( \cos\frac{\theta}{2} |\!\uparrow\downarrow\downarrow\rangle + \sin\frac{\theta}{2} |\!\downarrow\downarrow\uparrow\rangle \right) ~,
\end{equation}
where $\theta$ is defined in the range $0 \le \theta \le \pi$ by
\begin{eqnarray}
\cos\theta &=& - \frac{\frac{\widetilde{J}}{2}+\widetilde{B}_c}{A_{-1/2}} ~,\\
\sin\theta &=& \frac{\sqrt{2}\widetilde{J}}{A_{-1/2}} ~.
\end{eqnarray}

We can now compute the concurrence between the probe qubits.
We first construct the eight-dimensional density matrix for the ground state, including the degrees of freedom for the two qubits and the pseudo-spin. 
We then trace out the pseudospin degree of freedom to obtain the reduced density matrix for qubits 1 and 2, which is given by
\begin{eqnarray} 
\rho_{12}
&=& \frac{1}{2}\cos^2\frac{\theta}{2} \left(  |\!\uparrow\downarrow \rangle 
+ |\!\downarrow\uparrow \rangle \right) \left( \langle\uparrow\downarrow\! | +  \langle\downarrow\uparrow\! |\right)
\nonumber \\ && 
+ \sin^2\frac{\theta}{2} |\!\downarrow\downarrow \rangle \langle\downarrow\downarrow\! |   \nonumber\\
&=& \label{eq:rho}
\left( \begin{array}{cccc}
           0 & 0 & 0 & 0 \\
           0 & \frac{1}{2}\cos^2\frac{\theta}{2} & \frac{1}{2}\cos^2\frac{\theta}{2} & 0 \\
           0 & \frac{1}{2}\cos^2\frac{\theta}{2} & \frac{1}{2}\cos^2\frac{\theta}{2} & 0 \\
           0 & 0 & 0 & \sin^2\frac{\theta}{2}
           \end{array}  
    \right) ~.                                   
\end{eqnarray}
From Eq.~(\ref{eq:concur}), we then obtain
\begin{equation}
C = \cos^2\frac{\theta}{2} 
= \frac{1}{2} \left( 1 - \frac{\frac{\widetilde{J}}{2}+\widetilde{B}_c}{\sqrt{\frac{9}{4}\widetilde{J}^2 + \widetilde{B}_c^2 + \widetilde{J} \widetilde{B}_c}} \right) ~.
\end{equation}
We see that the concurrence attains a maximum value of 1/3 at the critical point,~\cite{oh_prb2010} and decreases to zero as $(J_c/B_c)^2$, when $B_c\gtrsim J_c$.
The full-width-at-half-max of the peak is given by $ (-1+\sqrt{32/5})\widetilde{J}  \simeq 1.53 \widetilde{J}$. 

We close this section by outlining an experimental procedure to observe concurrence peaks associated with QPTs in a spin chain.
The spin system is prepared in its ground state, for example by thermalization or an adiabatic initialization procedure.
Since the probe qubits are nonproximal, the concurrence measurement requires performing quantum state tomography of the two-qubit reduced density matrix. 
The full version of this technique involves 15 separate measurements of the two-qubit correlators $\{\sigma_{1i}\sigma_{2j}\}$, where $\sigma_{1i}$ ($\sigma_{2j}$) is a Pauli operator acting on qubit 1 (2), with $i,j \in \{I,X,Y,Z \}$. \cite{nielsen_chuang_book}
(We exclude the trivial two-qubit identity operator.)

\section{Local probes}\label{sec:4_local}
The local probe geometry that we consider is shown in Fig.~\ref{fig:1_system}(d).
Here, one side of a double-quantum dot is attached to one node of the spin chain.
The probe Hamiltonian is then given by
\begin{equation}
H_p=J_{12} \mathbf{S}_1 \cdot \mathbf{S}_2 + J_{2c} \mathbf{S}_2 \cdot \mathbf{s}_1 ~.
\end{equation}

\begin{figure}
  \includegraphics[width=\linewidth]{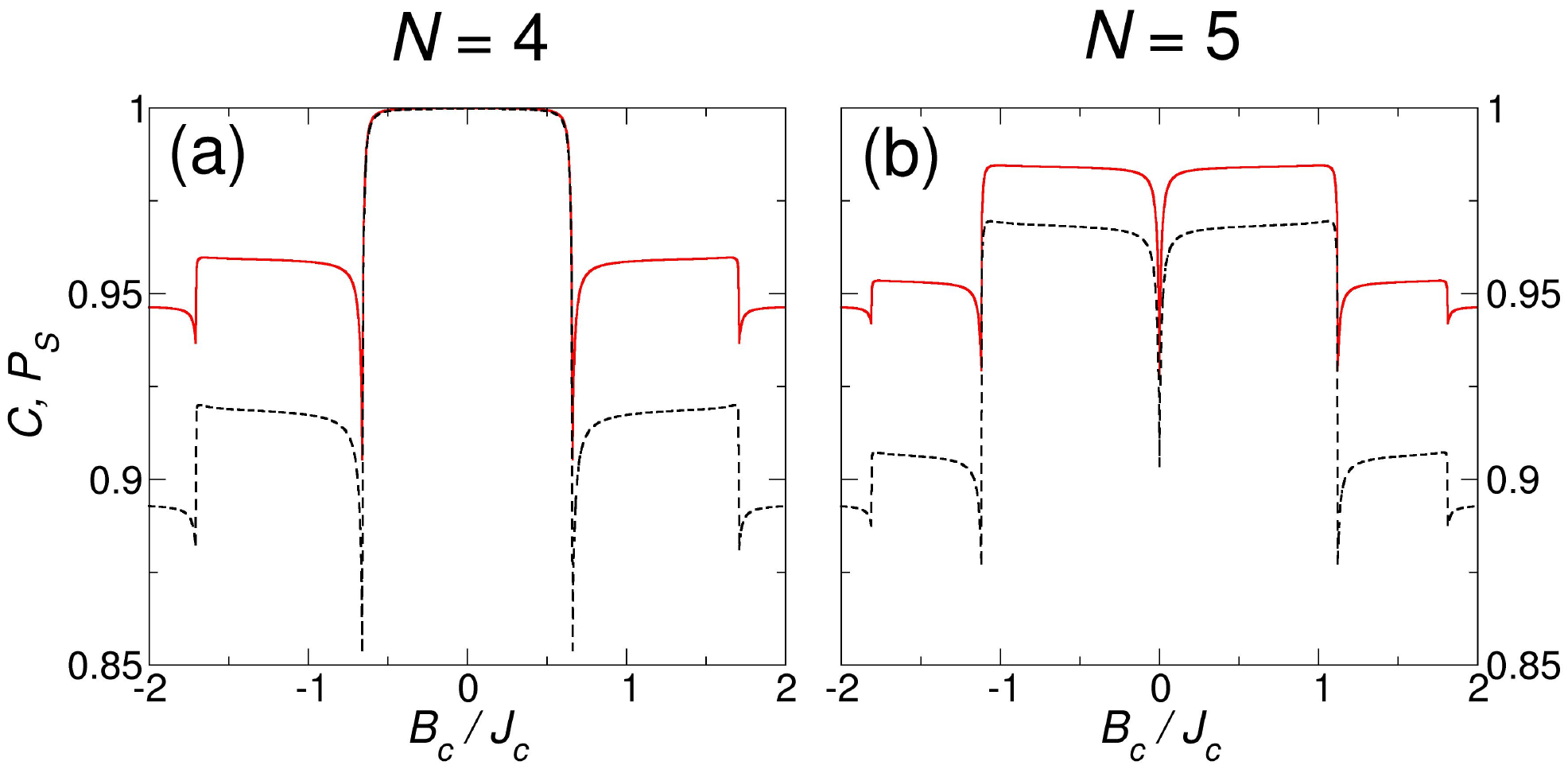}\\
  \caption{(color online) The concurrence $C$ and singlet probability $P_S$ (dashed black and solid red curves, respectively) of a double-dot, local probe coupled to a spin chain of size
          (a) $N=4$ and (b) $N=5$, as the chain undergoes QPTs. 
          Here we take $J_{12}$=$J_{2c}$=$0.02J_c$.
          $C$ and $P_S$ are both dimensionless.                  
          }
  \label{fig:3_N04N05}
\end{figure}

Using the methods described above, we can compute the concurrence between the spins in the probe double dot, obtaining the results shown in Fig.~\ref{fig:3_N04N05} for two different size chains.
Similar to the nonlocal probe, the QPTs in the spin chain are imprinted onto the local probe in the form of concurrence singularities.
In this case, the singularities are downward-pointing, rather than upward-pointing.

In Fig.~\ref{fig:3_N04N05}, we also show the overlap probability between the probe qubits and a singlet state.
Similar to the concurrence, the singlet probability $P_S$ is determined by first obtaining the reduced density matrix for the two probe qubits $\rho_{12}$, then computing
\begin{equation}
P_S=\langle S|\rho_{12}|S\rangle ,
\end{equation} 
where $|S\rangle =(|\!\uparrow\downarrow\rangle-|\!\downarrow\uparrow\rangle)/\sqrt{2}$ is the usual singlet state.
In Fig.~\ref{fig:3_N04N05}, we see that singularities also arise in $P_S$, which closely mirror those in $C$.

We can understand the main features in Fig.~\ref{fig:3_N04N05} by applying perturbation theory to the local probe geometry.
(Again, details are provided in Appendix~\ref{sec:A_eff_param}.)
As before, we note that the chain is effectively inert away from a critical point.
In this case, the noncritical effective Hamiltonian is given by
\begin{equation}\label{eq:H_l_nc}
\widetilde{H}_\text{nc} = \widetilde{B}_2 S_{2,z} + J_{12} \mathbf{S}_1 \cdot \mathbf{S}_2 ~.
\end{equation}
The entanglement between the probe qubits is determined by the interplay between $J_{12}$ which enhances the concurrence, and $\widetilde{B}_2$ which suppresses it.
The effective local field, $\widetilde{B}_2=\langle 0|s_{1z}|0 \rangle$, depends only on the true spin of the ground state of the spin chain, $|0\rangle$. 
In contrast with the pseudospin, which remains fixed at 0, the true spin increases by 1 in each successive noncritical region, as we move away from the value $B_c=0$.
Accordingly, the concurrence is suppressed in discrete steps.

Approaching a critical point, the chain becomes pseudospin-1/2, and the effective Hamiltonian takes the form
\begin{equation}\label{eq:H_l_c}
\widetilde{H}_\text{cp} = \widetilde{B}_2 S_{2,z} - \widetilde{B}_c S_{cz}
+\sum_{\alpha =x,y,z} \widetilde{J}_{2c\alpha} S_{2\alpha} S_{c\alpha} ~.
\end{equation}
The singularities all have a downward-pointing ``valley" shape, which can be explained by considering the $B_c$=0 transition in Fig.~\ref{fig:3_N04N05}(b). 
In the non-critical region, $\widetilde{B}_2\neq 0$, while in the critical region, $\widetilde{B}_2$=0. 
(The latter is only true for the $B_c$=0 transition.)  
This would normally lead to an upward-pointing singularity, since $\widetilde{B}_2$ suppresses the entanglement.
However, a second effective coupling ($\widetilde{J}_{2c}$) emerges in the critical region, as indicated in Eq.~(\ref{eq:H_l_c}), which reduces the entanglement between qubits 1 and 2 by sharing the entanglement with pseudospin $c$.
Below, we show that the latter effect is always dominant, leading to valley-type singularities.

\begin{figure}
  \includegraphics[width=1.7in]{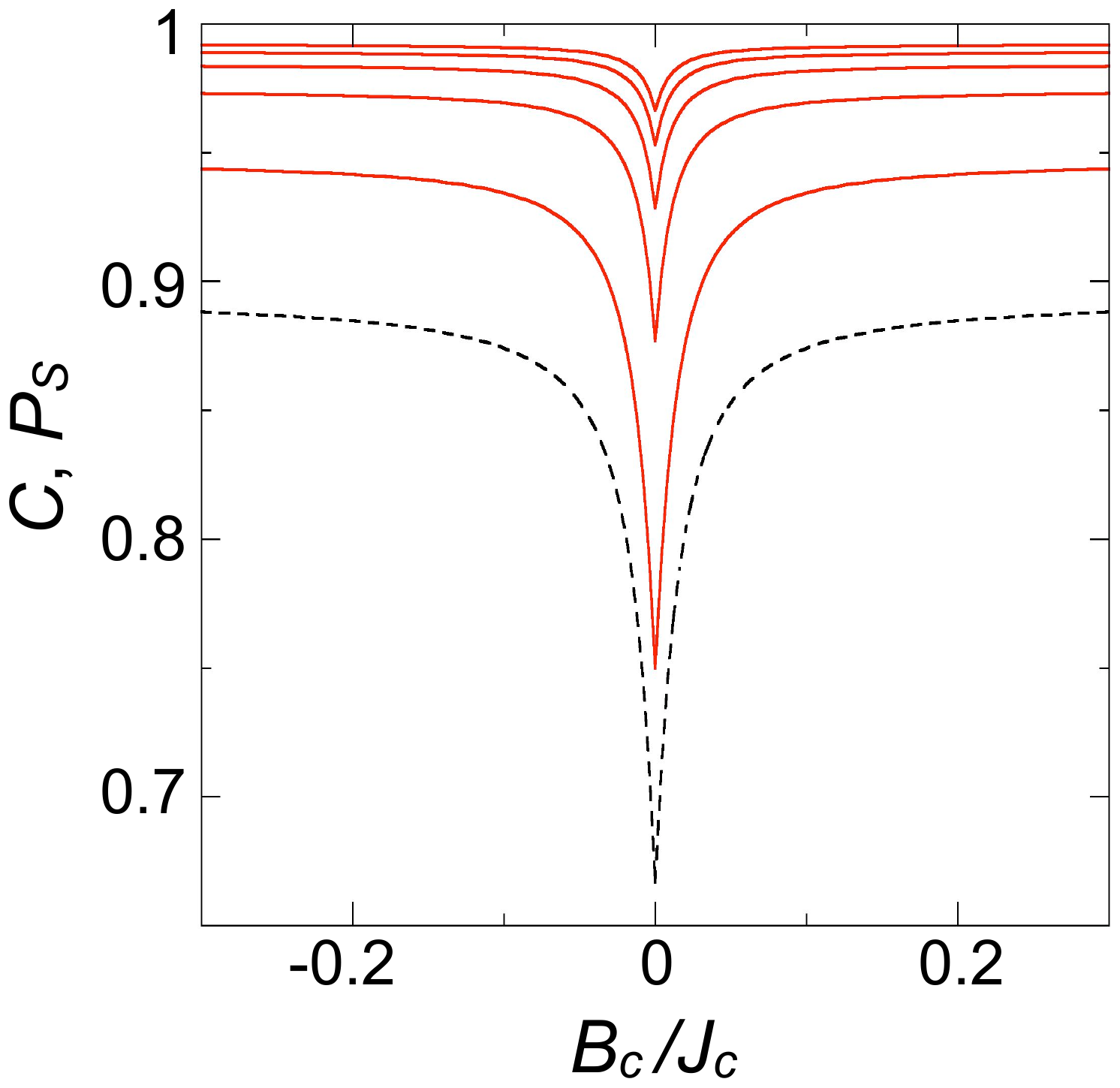}\\
  \caption{(color online) 
  Solid red curves show the singlet probabilities $P_S$ for a double quantum dot coupled locally to a spin chain of length $N$, as a function of the magnetic field on the chain.
  From top to bottom, the curves correspond to $N=9,7,5,3,$ and 1.
  The dashed black curve shows the corresponding concurrence $C$ for the case $N=1$.
  Here, we take $J_{12}$=$J_{2c}$=$0.02J_c$.
  (For the case $N=1$, $J_c$ is simply a reference value.)
          }
  \label{fig:4_N01}
\end{figure}

We obtain exact, analytical solutions, focusing strictly on the $B_c$=0 transitions in odd-size spin chains.
We obtain results in three different regimes:  
(i) the special point $B_c=0$, 
(ii) the critical regime about this QPT, 
and (iii) the asymptotic, noncritical regime between $B_c=0$ and any nearby critical points.
Numerical results over the entire range are shown in Fig.~\ref{fig:4_N01}, for chains of varying length.

(i) At the special point $B_c$=0, we find that $\widetilde{B}_c$=$\widetilde{B}_2$=$0$, and the effective coupling $\widetilde{J}_{2c}$ is isotropic, yielding the effective Hamiltonian
\begin{equation}\label{eq:H_l_c_B0}
\widetilde{H}_\text{cp}=(J_{12}{\bf S}_1+\widetilde{J}_{2c}{\bf S}_c)\cdot{\bf S}_2~.
\end{equation}
Since $C$ and $P_S$ are dimensionless quantities, at the point $B_c$=$0$ they can only be functions of the dimensionless ratio $\gamma = \widetilde{J}_{2c}/J_{12}$.
Since $S_{\text{tot}}^2=(\mathbf{S}_1+\mathbf{S}_2+\mathbf{S}_c)^2$ and  
$S_{\text{tot},z}=S_{1,z}+S_{2,z}+S_{c,z}$ commute with the effective Hamiltonian, 
similar to the nonlocal probe case, we determine the ground state of Eq.~(\ref{eq:H_l_c_B0}) by solving it in each of the subspaces labelled by $S_{\text{tot}}$ and $S_{\text{tot},z}$. 
In this way, we obtain the ground state 
\begin{eqnarray}
|\Psi_{\text{GS}}\rangle &=& \left( \frac{1}{\sqrt{2}} \cos\frac{\theta}{2} + \frac{1}{\sqrt{6}} \sin\frac{\theta}{2} \right)  \big( |\!\uparrow\downarrow\uparrow\rangle -|\!\downarrow\uparrow\uparrow\rangle \big)   \nonumber\\
 &&  -\sqrt{\frac{2}{3}}\sin\frac{\theta}{2} |\!\uparrow\uparrow\downarrow \rangle ~,
\end{eqnarray}
where $\theta$ is defined in the range $0 \le \theta \le \pi$ by
\begin{eqnarray}
\cos\theta &=& \frac{J_{12}-\frac{\widetilde{J}_{2c}}{2}}{\sqrt{J_{12}^2-J_{12}\widetilde{J}_{2c}+\widetilde{J}_{2c}^2}} ~,\\
\sin\theta &=& \frac{\frac{\sqrt{3}}{2}\widetilde{J}_{2c}}{\sqrt{J_{12}^2-J_{12}\widetilde{J}_{2c}+\widetilde{J}_{2c}^2}} ~.
\end{eqnarray}

We can compute the concurrence and singlet probability for the two probe qubits.
At the special point $B_c=0$, they are given by
\begin{gather}
C = \frac{1}{3} \left( 1 + \frac{2-\gamma}{\sqrt{1-\gamma+\gamma^2}} \right)~. \label{eq:C_l_c} \\
P_S = \frac{1 + 3 C}{4} ~.\label{eq:PS_c}
\end{gather}

(ii) It is possible to obtain exact results for $C$ and $P_S$ in the critical regime about $B_c=0$, when $|B_c|>0$.
The calculation is tedious but straightforward, as described in Appendix~\ref{sec:B_concur_local}.  
The result shows that the essential singularity for both $C$ and $P_S$ is linear, as consistent with Fig.~\ref{fig:4_N01}.
The analytical expressions for the slopes, on either side of $B_c=0$, are rather complicated however.

(iii) Far away from a critical point, Eq.~(\ref{eq:H_l_nc}) is valid, so $C$ and $P_S$ can only be functions of the dimensionless ratio $\widetilde{B}_{2}/J_{12}$.
In this case, $S_{\text{tot},z}$ commutes with the effective Hamiltonian, 
and we readily obtain the two eigenstates $|\uparrow\uparrow\rangle,|\downarrow\downarrow\rangle$, 
corresponding to $S_{\text{tot},z}=\pm 1$ with eigenvalues $E_{\pm 1}=J_{12}/4 \pm \widetilde{B}_2/2$.
In the subspace of $S_{\text{tot},z}=0$, the Hamiltonian matrix in the basis  $\{|\uparrow\downarrow\rangle,|\downarrow\uparrow\rangle \}$ is given by
\begin{equation}
H_0 = \left( \begin{array}{cc}
           \frac{-J_{12}}{4} - \frac{\widetilde{B}_2}{2}  & \frac{J_{12}}{2}  \\
           \frac{J_{12}}{2} &  \frac{-J_{12}}{4} + \frac{\widetilde{B}_2}{2}
           \end{array}
    \right)~.
\end{equation}
The lowest eigenvalue in this subspace is $E_0=-J_{12}/4 - \sqrt{J_{12}^2+\widetilde{B}_2^2}\Big/ 2$ and the eigenstate is 
$|\Psi_0\rangle = \cos \frac{\theta}{2} |\uparrow\downarrow\rangle - \sin \frac{\theta}{2} |\downarrow\uparrow\rangle$, 
where $\theta$ is defined in the range $0 \le \theta \le \pi$ by
\begin{eqnarray}
\cos\theta &=& \sqrt{ \frac{1+\widetilde{B}_2/\sqrt{J_{12}^2+\widetilde{B}_2^2}}{2} } ~,\\
\sin\theta &=& \sqrt{ \frac{1-\widetilde{B}_2/\sqrt{J_{12}^2+\widetilde{B}_2^2}}{2} } ~.
\end{eqnarray}
We can easily verify that $|\Psi_0\rangle$ is the ground state for any value of $\widetilde{B}_2$ 

We can now obtain expressions for the concurrence and singlet probability.
In Appendix \ref{sec:A_eff_param}, we show that $\widetilde{B}_{2}/J_{12}=\widetilde{J}_{2c}/2J_{12}=\gamma/2$ for the $B_c=0$ transition, which leads to simplifications in the expressions:
\begin{eqnarray}
C &=&  \frac{1}{\sqrt{1+\gamma^2/4}} ~, \label{eq:C_l_nc}\\
P_S &=& \frac{1+C}{2} ~. \label{eq:PS_nc}
\end{eqnarray}
In the limit of large chain size, $N \gg 1$, we have previously shown~\cite{friesen_biswas_prl2007,oh_prb2012} that $\gamma\sim N^{-1/2} \rightarrow 0$.
Hence, $C$ and $P_S$ approach the constant value of 1.

To conclude this section, we note that the results described above can be used to compute the depth of the valleys in $C$ and $P_S$.
Here, we define the valley depth as the difference between the asymptotic limits, for large and small values of $|B_c|$.
We again consider the limit $N\gg 1$ and $\gamma\rightarrow 0$ in Eqs.~(\ref{eq:C_l_c}), (\ref{eq:PS_c}), (\ref{eq:C_l_nc}), and (\ref{eq:PS_nc}), finding that the $C$ and $P_S$ valley depths approach zero quickly, as $\gamma^2/8\sim 1/N$. 
Hence, the singularity is suppressed, as consistent with the numerical results shown in Fig.~\ref{fig:4_N01}.

\begin{figure}
  \includegraphics[width=\linewidth]{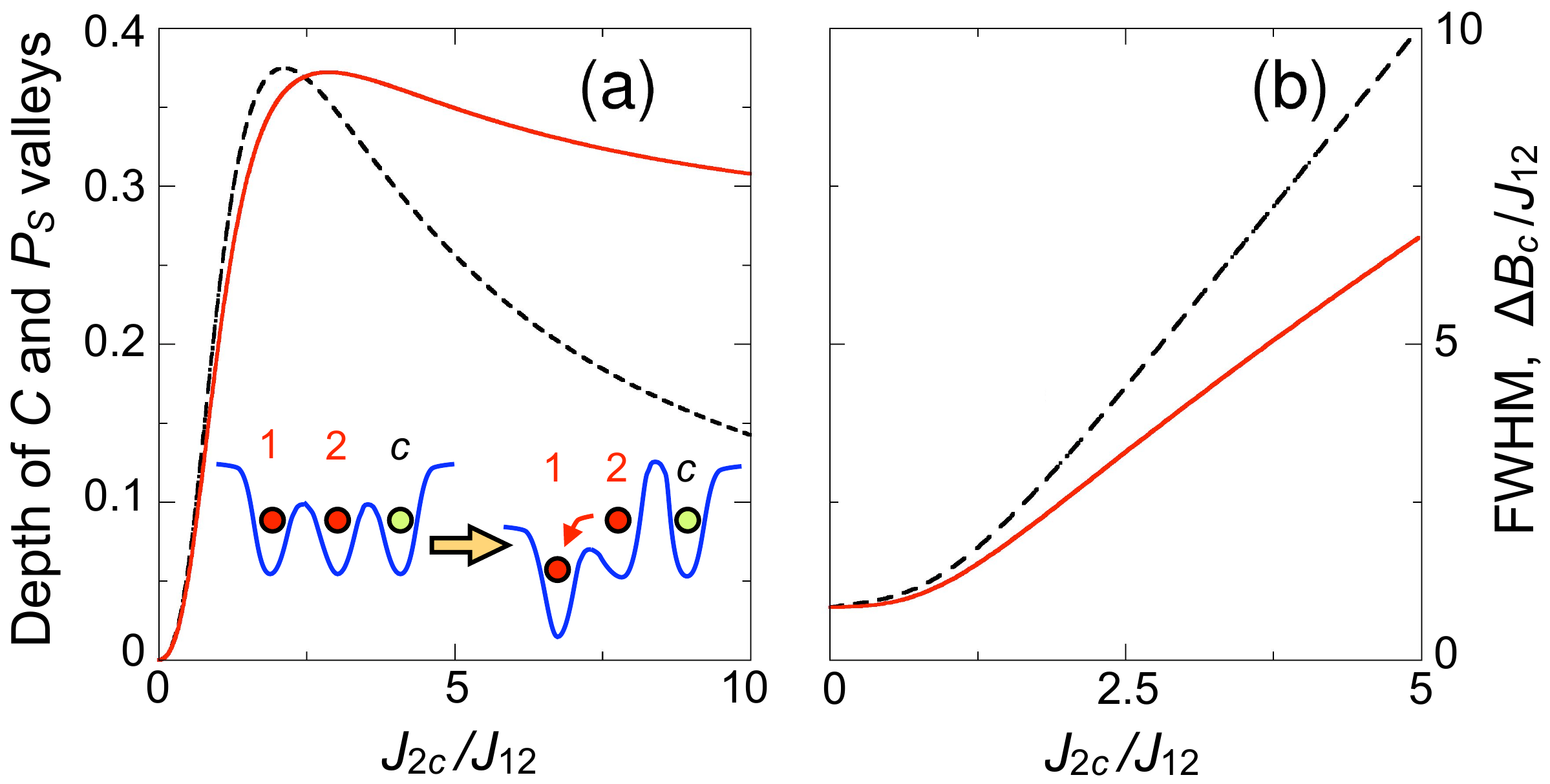}\\
  \caption{(color online) (a) Depth of $C$ and $P_S$ valley singularities, like those shown in Fig.~\ref{fig:4_N01} (with $N=1$), when two probe qubits (1 and 2) are coupled to a quantum dot ($c$), plotted as a function of the coupling ratio. 
  The inset shows the confinement profile of the triple-dot experiment.
           (b) Full-width-at-half-minimum (FWHM) of the valleys with $J_{12}$ fixed at 0.02$J_c$ .
           In both panels, the concurrence $C$ and the singlet probability $P_S$ are represented by dashed black and solid red curves, respectively.
           } 
  \label{fig:5_N01_changeJ}
\end{figure}

\section{Triple-dot experiment}\label{sec:5_experiment}
We now propose an experiment to investigate the QPT in the opposite limit, $N=1$.
We consider the triple quantum dot geometry shown in the inset of Fig.~\ref{fig:5_N01_changeJ}(a).
For simplicity, we assume the dots are singly occupied.
The double quantum dot (on the left) is used to probe a spin ``chain" of length 1 (on the right), whose ground state properties change dramatically as a function of the applied field $B_c$ at the transition point $B_c=0$.
The field is applied only to spin $c$, necessitating a gradient scheme to cancel out the field on the probe dots.
(See further discussion of this point below.)
The experiment proceeds by first preparing the triple dot in its ground state.
We then turn off the exchange coupling to dot $c$ and detune the probe double dot, so that the electron in dot~2 moves to dot~1 only when the probe is in a singlet state, due to the large singlet-triplet energy splitting for two electrons in a single dot.~\cite{Petta2005}  
The singlet probability can then be monitored using charge sensing, via a nearby charge sensor.

For the case $N=1$, it can be shown that $\widetilde{B}_2 = \pm J_{2c}/2$ in Eq.~(\ref{eq:H_l_nc}).
We can then determine the depth of the $C$ and $P_S$ valleys, as well as their widths, as a function of the ratio $\gamma=J_{2c}/J_{12}$, as shown in  Fig.~\ref{fig:5_N01_changeJ}.
Generally, we see that applied fields of order $B_c\sim J_{12}$ are needed, to observe the QPT.  
Since quantum dot exchange couplings are typically of order $\mu$eV, the magnetic field differences between the quantum dots needed in this experiment are of order 10~mT, which is relatively easy to achieve in the laboratory.~\cite{pioro_np2008}

\begin{figure}
  \includegraphics[width=\linewidth]{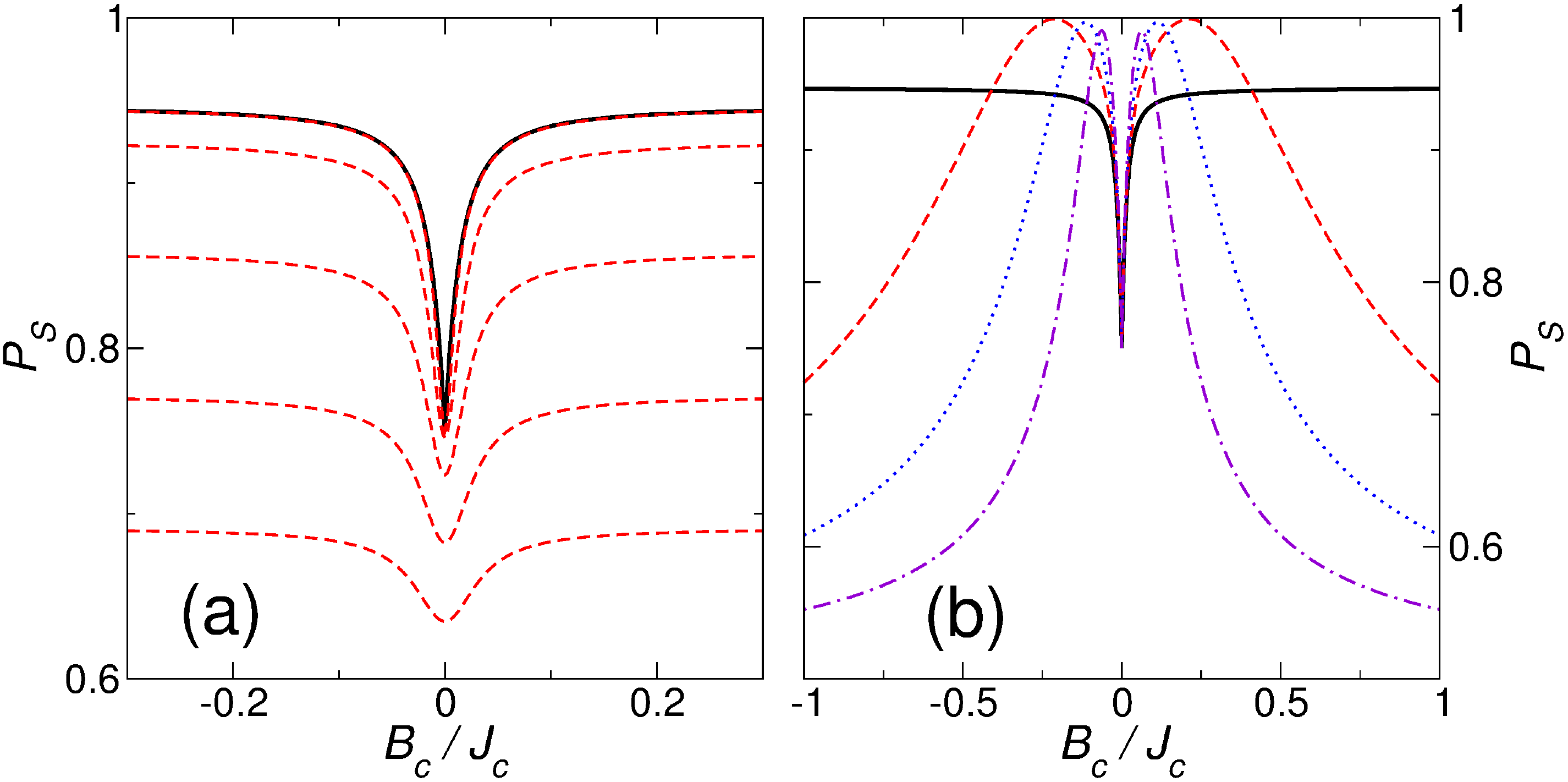}\\
  \caption{(color online)   
  Variations of the singlet probability, for a spin chain of size $N=1$, with $J_{12}$=$J_{2c}$=0.02$J_c$
  (a) Variations due to nonzero temperature.
  The solid black curve corresponds to $T$=0, while from top to bottom, the dashed red curves correspond to $k_B T/J_{12}$=0.1, 0.2, 0.3, 0.4, and 0.5 .
  (b) Variations due to the presence of an unwanted magnetic field, $B_2$=$\delta_B B_c$.
  The curves correspond to $\delta_B$=0 (solid black), 0.05 (dashed red), 0.1 (dotted blue), and 0.2 (dot-dashed purple).           
  } 
  \label{fig:6_disorder}
\end{figure}

The proposed triple-dot QPT experiment will be affected by decoherence.
However, after initializing the ground state, the protocol involves just two quick operations:  turning off the interaction between dots 2 and $c$, and performing the singlet projection.
Both of these operations have been performed successfully in double dot experiments.\cite{Petta2005} 
The same is expected to be true here, and we do not further explore effects directly related to decoherence.

Thermal effects play another role however, through a mechanism unrelated to decoherence.
During the initialization process, the triple dot is supposed to relax into its ground state.
At non-zero temperatures, $k_B T \lesssim J_{12},J_{2c}$, there will be a population of excited spin states, in addition to the ground state.
Since the QPT is associated specifically with the ground state, the net effect is to wash out singularity.

To explore this effect, we can introduce a thermal population of initial states of the three-spin Hamiltonian.
For the full set of eigenstates $\{|\Psi_\alpha \rangle\}$, with the corresponding eigenvalues $E_\alpha$, the distribution probabilities are given by
\begin{equation}
P_\alpha = e^{-E_\alpha/k_BT}/Z.
\end{equation}
Here,
\begin{equation}
Z=\sum_\alpha e^{-E_\alpha /k_BT} 
\end{equation}
is the classical partition function.
The initial mixed state is then described by the density matrix
\begin{equation}
\rho =\sum_\alpha P_\alpha |\Psi_\alpha \rangle \langle \Psi_\alpha | .
\end{equation}

Proceeding as usual, we compute the reduced density matrix for the probe qubits, $\rho_{12}$, and calculate the singlet projection $P_S$.
The results are shown in Fig.~\ref{fig:6_disorder}(a).
At low temperatures, the singlet composition is nearly the same as zero temperature.
As the temperature increases, the asymptotic  (large $|B_c|$) value of the singlet probability decreases, and the depth of the valley is reduced.
The QPT singularity is still evident for temperatures of order of the exchange interactions,  $k_BT \sim J_{12}, J_{2c}$. 
At high temperatures however, $P_S$ approaches the limiting value of 1/4, and the valley disappears. 

Other types of control errors can arise in the proposed triple dot experiment.
For example, it may be difficult to tailor the magnetic fields such that $B_c$ is tunable, while $B_1$ and $B_2$ remain constant, and equal to zero.
For typical quantum dots, the exchange interactions are of order $J_{12},J_{2c} \sim 1$-10~$\mu$eV;\cite{Petta2005,maune_borselli_nature2012}
the experiment therefore requires magnetic fields of order $B_c\sim 10$~mT.
Field differences greater than 10~mT can be attained on closely space quantum dots,\cite{pioro_np2008} so the proposed range of magnetic fields is not unreasonable.  
If we incorporate two current-carrying wires near the spin chain, it should be possible to tune $B_c$ while keeping $B_1$=0.
However, it is more challenging to simultaneously enforce zero field on both quantum dots, 1 and 2.

We now repeat our numerical calculations of $P_S$, assuming that $B_c$ is tunable, while $B_1=0$ and $B_2=\delta_B B_c$.
In other words, we assume that an unwanted field arises on qubit 2 due to magnetic crosstalk, with a proportionality constant $\delta_B<1$ that is fixed, and depends only on the experimental geometry.
The results of our calculation are shown in Fig.~\ref{fig:6_disorder}(b).
Several changes in behavior are observed when $B_2\neq 0$.
First, we see that the singlet probability is nearly unaffected when $B_c$ is very small.
It then peaks at a value of 1, and finally falls below the (solid black) $B_2=0$ curve at higher fields. 

The fact that the valley in $P_S$ is nearly unaffected for very small $B_c$ values is easy to understand.
In this range, $B_2$ is also very small, so the exchange interaction is the dominant term for determining the ground state.
The same is true when $B_2=0$; hence $P_S$ is unaffected.

The fact that $P_S$ peaks at 1 is also easy to understand.
We have previously noted that qubit 2 experiences an effective magnetic field $\widetilde{B}_2$ due to the exchange interaction. $\widetilde{B}_2$=0 close to the critical point $B_c$=0, but away from the critical point,
this effective field points in the opposite direction of the physical field.
For some value of $B_c$, the two fields can cancel each other.
This cancellation occurs in the noncritical regime unless $\delta_B$ is too large.
The effective system is then simply exchange-coupled double-dot probe in effectively zero magnetic field. 
Therefore the ground state corresponds precisely to the singlet state of the probe.
As consistent with Fig. \ref{fig:6_disorder}(b), the value of $B_c$ needed for this cancellation decreases, for larger values of $\delta_B$. 
Since the maximum value of $P_S$ increases to 1 in this scenario, it seems that small local variations in the magnetic field could actually make it easier to observe a QPT.

\section{Summary and conclusions}\label{sec:6_summary}
In summary, we have shown that QPTs in a quantum dot spin chain cause non-analytic behavior in the entanglement between external qubits, in the form of local or nonlocal probes, and we have argued that there are important practical benefits for using a local probe.
Based on our analysis of the local probe, we have proposed an experiment to observe a QPT in a single quantum dot spin, where the ground state changes character abruptly as a function of magnetic field.  
The experiment uses a triple quantum dot, and we have shown that the requirements are realistic, even taking into account finite temperature and local field fluctuations.

We have found that QPTs can be observed in other types of spin chains (\emph{e.g.}, $XY$) using a local probe, although we do not describe those results here.
We therefore believe that the local probe could be very useful for investigating entanglement in many quantum dot geometries, because of the practicality of performing singlet projection measurements.

This work was supported by the DARPA QuEST program through a grant from AFOSR, and by NSA/LPS through grants from ARO (W911NF-08-1-0482 and W911NF-09-1-0393).

\begin{appendix}
\section{Expressions for effective couplings and local fields}\label{sec:A_eff_param}
In this appendix, we provide the coupling constants for four different effective Hamiltonians, Eqs. (\ref{eq:H_nl_nc}), (\ref{eq:H_nl_c}), (\ref{eq:H_l_nc}), and (\ref{eq:H_l_c}), corresponding to local or nonlocal probes coupled to a spin chain. 
In the critical regime, the spin chain is represented as pseudospin-1/2, while
in the noncritical regime, it is represented as pseudospin-0.
The following formulae were first derived in Ref.~[\onlinecite{shim_oh_prl2011}].

For the nonlocal probe in the noncritical regime [Eq.~(\ref{eq:H_nl_nc})]:
\begin{gather}
\widetilde{B}_{1} = J_{1c} \langle 0 | s_{1z} | 0 \rangle ~,\\
\widetilde{B}_{2} = J_{2c} \langle 0 | s_{Nz} | 0 \rangle ~,\\
\widetilde{J}_{12\alpha} = -2\sum_{m>0} \frac{J_{1c} J_{2c}}{\vep_m-\vep_0} \langle 0 | s_{1\alpha} | m \rangle \langle m | s_{N\alpha} | 0 \rangle  ~, 
\end{gather}
for $\alpha$=$x,y,z$.
Here $|m \rangle$ corresponds to the $m$-th eigenstate of the spin chain Hamiltonian, $H_c$, with energy $\vep_m$.
$m=0$ corresponds to the unique ground state.

For the nonlocal probe in the critical regime [Eq.~(\ref{eq:H_nl_c})], we define the pseudospin ``up" state ($\ru$) and "down" state ($\rd$) as the ground states of the spin chain on either side of the transition, for higher and lower magnetic fields, respectively. 
The effective variables are
\begin{gather}
\widetilde{B}_{1} = \frac{J_{1c}}{2} \left( \lu s_{1z} \ru  + \ld s_{1z} \rd \right) ~, \\
\widetilde{B}_{2} = \frac{J_{2c}}{2} \left( \lu s_{Nz} \ru  + \ld s_{Nz} \rd \right) ~, \\
\widetilde{B}_c = \vep_{\Downarrow} - \vep_{\Uparrow} ~, \\
\widetilde{J}_{1c\alpha} = 
\left\{ \begin{array}{lc}
        J_{1c} \lu s_{1+} \rd  & \hspace{-.1in} (\alpha=x,y)   \\
        J_{1c} \left( \lu s_{1z} \ru - \ld | s_{1z} \rd  \right)  & (\alpha=z) 
        \end{array}
\right.   ~, \\
\widetilde{J}_{2c\alpha} = 
\left\{ \begin{array}{lc}
        J_{2c} \lu s_{N+} \rd  & \hspace{-.1in} (\alpha=x,y)   \\
        J_{2c} \left( \lu s_{Nz} \ru - \ld | s_{Nz} \rd  \right)  & (\alpha=z) 
        \end{array}
\right.  .
\end{gather}

For the local probe in the noncritical regime [Eq.~(\ref{eq:H_l_nc})]
\begin{eqnarray}
\widetilde{B}_{2} &=& J_{2c} \langle 0 | s_{1z} | 0 \rangle ~.
\end{eqnarray}

For the local probe in the critical regime [Eq.~(\ref{eq:H_l_c})]
\begin{gather}
\widetilde{B}_{2} = \frac{J_{2c}}{2} \left( \lu s_{1z} \ru  + \ld s_{1z} \rd \right) ~, \\
\widetilde{B}_c = \vep_{\Downarrow} - \vep_{\Uparrow} ~, \\
\widetilde{J}_{2c\alpha} =  
\left\{ \begin{array}{lc}
        J_{2c} \lu s_{1+} \rd & \hspace{-.1in} (\alpha=x,y) \\
        J_{2c} \left( \lu s_{1z} \ru - \ld | s_{1z} \rd  \right) & (\alpha=z)
        \end{array}
\right. .
\end{gather}

\section{Concurrence and singlet probability of a local probe connected to an odd-size spin chain}\label{sec:B_concur_local}

In this appendix, we derive general expressions for the concurrence and singlet probability of a local probe coupled to a pseudospin in the critical regime of the $B_c=0$ transition.
Equations~(\ref{eq:C_l_c}) and (\ref{eq:PS_c}) are special cases of this result.

For this particular QPT critical point, the effective Hamiltonian is isotropic, and Eq.~(\ref{eq:H_l_c}) reduces to
\begin{equation}
\widetilde{H}_\text{cp} = J_{12} \mathbf{S}_1 \cdot \mathbf{S}_2 + \widetilde{J}_{2c} \mathbf{S}_2 \cdot \mathbf{S}_c - B_c S_{c,z}  ~.
\end{equation}
The spin operator $S_{\text{tot},z}=S_{1,z}+S_{2,z}+S_{c,z}$ commutes with the Hamiltonian, so we may divide the full Hilbert space into subspaces according to their $S_{\text{tot},z}$ labels. 
For $S_{\text{tot},z}=\pm 3/2$, there are two one-dimensional subspaces with eigenstates $| S_{1,z}, S_{2,z}, S_{c,z} \rangle=| \uparrow \uparrow \uparrow \rangle$ and $| \downarrow \downarrow \downarrow \rangle$. 
The eigenenergies are $E_{\pm 3/2} = (J_{12}+\widetilde{J}_{2c})/4 \mp B_c/2$, respectively. 
For $S_{\text{tot},z}=1/2$, the subspace is three dimensional and the Hamiltonian matrix in the ordered basis $\{ |\uparrow\uparrow\downarrow\rangle, |\uparrow\downarrow\uparrow\rangle, |\downarrow\uparrow\uparrow\rangle \}$ is given by
\begin{equation}
H = \left(  \begin{array}{ccc} 
            \frac{J_{12}-\widetilde{J}_{2c}+2B_c}{4}  &  \frac{\widetilde{J}_{2c}}{2}  &  0  \\
            \frac{\widetilde{J}_{2c}}{2}  &  \frac{-J_{12}-\widetilde{J}_{2c}-2B_c}{4} & \frac{J_{12}}{2} \\
            0  &  \frac{J_{12}}{2}  &  \frac{-J_{12}+\widetilde{J}_{2c}-2B_c}{4}
            \end{array} 
    \right)~.
\end{equation}
We can obtain the eigenvalues by solving the cubic function $| H - \lambda I |=0$.
The ground state energy in this subspace is
\begin{equation}
\lambda_1 = -\frac{1}{12} \left( J_{12}+\widetilde{J}_{2c}+2B_c + 2 \sqrt{D_1} \cos \frac{\phi}{3} \right)~,
\end{equation} 
where $\phi$ is defined in the range $0 \le \phi \le \pi$ by
\begin{eqnarray}
\cos \phi &=& \frac{D_2}{2 D_1^{3/2}} ~, \\
\sin \phi &=& \frac{\sqrt{27\Delta}}{2 D_1^{3/2}} ~,
\end{eqnarray}
and 
\begin{eqnarray}
D_1 &=& 4 \left( 4 J_{12}^2 + 4 \widetilde{J}_{2c}^2 + 4 B_c^2 - J_{12} \widetilde{J}_{2c} 
\right. \nonumber \\ && \hspace{.25in} \left.
+ 4 J_{12} B_c - 2 \widetilde{J}_{2c} B_c \right) ~,\\
D_2 &=& 16 \left( 8 J_{12}^3 + 8 \widetilde{J}_{2c}^3 - 8 B_c^3 - 3 J_{12}^2 \widetilde{J}_{2c} - 3 J_{12} \widetilde{J}_{2c}^2  
\right. \nonumber \\ && \hspace{.25in} \left.
+ 6 J_{12} \widetilde{J}_{2c} B_c -12 J_{12} B_c^2 + 6 \widetilde{J}_{2c} B_c^2 
\right. \nonumber \\ && \hspace{.25in} \left.
+ 12 J_{12}^2 B_c - 6 \widetilde{J}_{2c}^2 B_c \right) ~, \\    
\Delta &=& \frac{1}{27} \left( 4 D_1^3 - D_2^2  \right) ~.             
\end{eqnarray}

Without loss of generality, we will only consider $B_c >0$.
In this case, the $S_{\text{tot},z}=1/2$ subspace always has a lower energy than the subspaces with $S_{\text{tot},z}=\pm 3/2$.
The ground state energy is then $\lambda_1$, and the corresponding eigenstate of the three-spin system is
\begin{equation}
|\Psi_\text{GS}\rangle = z_1 |\uparrow\uparrow\downarrow\rangle + z_2 |\uparrow\downarrow\uparrow\rangle + z_3 |\downarrow\uparrow\uparrow\rangle ~,
\end{equation}
where
\begin{eqnarray}
z_1 &=& - \frac{\widetilde{J}_{2c}}{2\sqrt{Z}( F - \lambda_1 )} ~,\\
z_2 &=& \frac{1}{\sqrt{Z}} ~,\\
z_3 &=& \frac{J_{12}}{2\sqrt{Z}(F + \lambda_1)} ~,\\
Z &=& 1+\frac{ (\frac{\widetilde{J}_{2c}}{2})^2 \left( \lambda_1 + F \right)^2 + (\frac{J_{12}}{2})^2 \left( \lambda_1 - F \right)^2 }
           { \left( \lambda_1^2-F^2 \right)^2 } ,\\
F &=& \frac{J_{12}-\widetilde{J}_{2c}+2B_c}{4}  ~.           
\end{eqnarray}
As before, we construct the eight-dimensional density matrix for the ground state and trace out the pseudospin degree of freedom.  
The resulting reduced density matrix for the probe double dot is given by
\begin{eqnarray}
\rho_{12}  
     &=& \left( \begin{array}{cccc}
                z_1^2 & 0 & 0 & 0 \\
                0 & z_2^2 & z_2 z_3 & 0 \\
                0 & z_2 z_3 & z_3^2 & 0 \\
                0 & 0 & 0 & 0 
                \end{array}  
         \right) ~.
\end{eqnarray}

We then compute the concurrence, giving
\begin{eqnarray}
C &=& 2 |z_2 z_3| = \frac{J_{12}}{Z | F+\lambda_1 |}~. \label{eq:Cf}
\end{eqnarray}
The singlet probability $P_S$ is given by
\begin{eqnarray}
P_S &=& \langle S | \rho_{12} | S \rangle = \frac{1}{2} \left( z_2 - z_3 \right)^2 \nonumber \\
     &=&  \frac{1}{2Z} \left( 1- \frac{J_{12}}{2(F+\lambda_1)} \right)^2 ~. \label{eq:PSf}
\end{eqnarray}

Equations (\ref{eq:Cf}) and (\ref{eq:PSf}) represent general results for the QPT at $B_c=0$, in the critical regime with $B_c\geq 0$. 
We can then obtain the special case solutions of Eqs.~(\ref{eq:C_l_c}) and (\ref{eq:PS_c}) by setting $B_c=0$.
The algebra is rather complicated however, and the solutions are more easily obtained by solving the $B_c=0$ Hamiltonian, as described in Sec.~\ref{sec:4_local}.

\end{appendix}

\bibliographystyle{apsrev4-1}

\end{document}